\documentclass[prd,twocolumn,floatfix,amsmath,nofootinbib,amssymb,floatfix]{revtex4}
\usepackage{graphicx,color,dcolumn,booktabs,bm,multirow}
\usepackage{longtable,lscape}
\usepackage{txfonts}
\usepackage{overpic}
\usepackage{amssymb}
\usepackage{array}
\usepackage{indentfirst}
\usepackage{feynmf}   
\usepackage{slashed}  
\usepackage{cases}
\usepackage{color}
\usepackage{multirow}
\usepackage{epstopdf}
\usepackage[utf8]{inputenc}
\usepackage{graphicx,color,dcolumn,booktabs,bm}
\usepackage[colorlinks,
citecolor=blue,
anchorcolor=red,
menucolor=red,
linkcolor=red,
filecolor=red,
runcolor=red,
urlcolor=blue,
frenchlinks=red]{hyperref}

\def\stat{\mathrm{(stat.)}}
\def\syst{\mathrm{(syst.)}}

\begin{document}
	
	\title{$\Xi(1620)$ production in $K^- p$ scattering process}
	\author{Quan-Yun Guo$^{1}$}
	\author{Zi-Li Yue$^{1}$}
	\author{Dian-Yong Chen$^{1,2}$\footnote{Corresponding author}}\email{chendy@seu.edu.cn}
	\affiliation{$^1$ School of Physics, Southeast University, Nanjing 210094, People's Republic of China}
	\affiliation{$^2$ Lanzhou Center for Theoretical Physics, Lanzhou University, Lanzhou 730000, China}
	\date{\today}

	\begin{abstract}
In the present work, the production of $\Xi(1620)$ in the $K^- p$ scattering process is investigated by using an effective Lagrangian approach, where $\Xi(1620)$ is considered as a $\bar{K} \Lambda$ molecular state. Our estimations indicate that the cross sections for $K^-p\to K^+ \Xi(1620)^-$ are $(1.48 ^{+ 1.12}_{-0.69}) \ \mathrm{\mu b}$ at $P_K=2.8 \ \mathrm{GeV}$, where the uncertainties are resulted from the variation of the model parameter. As for the $K^-p\to K^+ \pi^0  \Xi^-$ process, the cross sections are estimated to be $(0.61 ^{+0.47}_{-0.29})\ \mathrm{\mu b}$ at $P_K =2.87 \ \mathrm{GeV}$, which is consistent with the experimental measurements.  
	\end{abstract}
	
	\maketitle

\section{Introduction}
\label{sec:Introduction}

Over the recent two decades, some multiquark candidates have been observed, and the majority of these states contains at least one heavy quark (see Refs. ~\cite{Guo:2017jvc,Olsen:2017bmm,Brambilla:2019esw,Ali:2017jda,Guo:2019twa,Liu:2013waa,Dong:2017gaw,Liu:2019zoy} for more details). The investigations of the heavy pentaquark system undoulbetly expanded our comprehension of the quark model. Such expansion of the quark model inherently predicts the multiquark states composed of only light quarks, whose masses are usually overlapped with the excited states of the traditional hadrons. Consequently, the identification of such kind of light exotic states experimentally or theoretically is rather difficult comparing to those with heavy quarks. Even so, some exotic candidates have been observed in the light sector, for example, the $Y(2175)$ could be considered as tetraquark state~\cite{Chen:2008ej, Wang:2006ri}, the $\Lambda(1405)$ is resulted from the $\bar{K} N$ interaction~\cite{Ezoe:2020piq,Oller:2000fj}, while the $\Omega(2012)$ is generated from the $\bar{H}\Xi^\ast $ interactions~\cite{Lu:2022puv,Liu:2020yen}. 

Besides the above examples, another states with $S=-2$, named $\Xi(1620)$ could also be a good exotic state candidate. The experimental observations of $\Xi(1620)$ could go back half a century. In Ref.~\cite{Apsell:1969sa}, the cross sections for the process $K^- p \to \Xi^- \pi^+ K^0$ had been measured at 2.87 GeV and significant structures was observed in the $\Xi^- \pi^+$ invariant mass spectrum at masses of 1630 MeV, the mass and width of $\Xi^0(1620)$ were reported to be $1628  \pm 5$ MeV and $15\pm 5$ MeV, respectively. Later, the cross sections for $K^- p \to \Xi^- \pi^+ K^0$ were also reported at 3.13, 3.30 and 3.58 GeV, and the mass and width of $\Xi^0(1620)$ were measured to be $1605.5 \pm 5.6$ MeV  and $ 20.8 \pm 7.4$ MeV. respectively~\cite{Ross:1972bf}. In addition, an investigations of $\Xi^\ast$ production in the $K^- p$ interactions were perform with the data at 2.87 GeV in in the $\Xi \pi$, $\Xi \pi \pi$, $\Xi(1530) \pi$, $Y\bar{K}$ and $Y\bar{K} \pi$ invariant mass distributions~\cite{Briefel:1977bp} and the cross sections for $K^- p \to \Xi(1620) K^0 \to \Xi^- \pi^+ K^0 $ was measured to be $(2.6 \pm 0.9)\ \mathrm{\mu b}$. After a long silence, the experimental breakthrough about $\Xi(1620)$ was appear in in 2019, when the Belle Collaboration announced their observation of $\Xi(1620)$ via the invariant mass spectrum of $\Xi^- \pi^+$ in the process $\Xi_c^+ \rightarrow \Xi^- \pi^+ \pi^+$~\cite{Belle:2018lws}. With the number of $\Xi(1620)^0$ events be 2 orders of magnitude larger than that in previous experiments, the mass and width were precisely determined to be
 \begin{eqnarray}
 	m&=& \left (1610.4 \pm 6.0 \stat ^{+6.1}_{-4.2} \syst \right) \ \mathrm{ MeV}, \nonumber\\
 	\Gamma &=&\left(59.9\pm 4.8 \stat^{+2.8}_{-7.1} \syst\right) \ \mathrm{MeV},
 \end{eqnarray}
 respectively.

On the theoretical side, it's worth mentioning that the observed mass of $\Lambda(1620)$ is very close to the threshold of $\Lambda \bar{K}$, which indicates that $\Xi(1620)$ could be considered as a $\Lambda \bar{K}$ molecular state. By using the unitary extension of the chiral perturbation theory, the low-energy meson-baryon scattering in the strangeness $S=-2$ sector was investigated, and a scattering-matrix pole was found around 1606 MeV with $J^P=1/2^-$, which could be identified with $\Xi(1620)$. The estimations with Bethe-Salpeter equations in Ref.~\cite{Wang:2019krq} indicated that the $\Xi(1620)$ could be interpreted as $\bar{K} \Lambda$ and $\bar{K} \Sigma$ bound states with $J^P=1/2^-$, and the decay properties of $\Xi(1620)$ in the $\bar{K} \Lambda-\bar{K}\Sigma$ molecule scenario also preferred the $J^P=1/2^-$ assignment~\cite{Huang:2020taj,Huang:2021ahp}. In the framework of the one-boson-exchange model, the authors in Ref.~\cite{Chen:2019uvv} suggested $\Xi(1620)$ as a $\bar{K} \Lambda$ molecule with $I(J^{P}) = 1/2 (1/2^-)$, which could be an analogue of $\Lambda(1405)$. However, there are different views of the inner structure of $\Xi(1620)$, for example, the analysis by using the chiral unitary model indicated that it is incompatible with the $K^-\Lambda$ scattering length extracted from the ALICE experimental data when considering $\Xi (1620)$ as a shallow quasi-bound state.  

So far the nature of $\Xi(1620)$ is still under debate, the investigations from different aspect are necessary to decoding its inner structure. Besides the mass spectrum and decay behavior, the production properties of $\Xi(1620)$ have also strong connection with its structure. It should be noted that the threshold of  $\bar{K} \Lambda$ and $K^- \Sigma^0$ are about $1610$ and $1686$ MeV, respectively. Considering the large discrepancy of the  $\bar{K} \Sigma$ threshold and $\Xi(1620)$ mass, in the present work, we consider $\Xi(1620)$ as a shallow bound state of $\bar{K} \Lambda$ with $J^P=1/2^-$, and investigate the production process $K^- p\rightarrow \Xi(1620)^- K^+$ by using the effective Lagrangian approach, where $\Xi (1620)$ could further decays into $\Xi(1314) \pi$.

This work is organized as follows. After introduction, we present the molecular structure of $\Xi(1620)$ and the cross sections for $K^- p \to K^+ \Xi(1620)^-$ and $K^- p \to K^+ (\Xi^- \pi^0)$.  In Section III,  the numerical results and related discussions of the cross sections are presented. The last section is devoted to a short summary.

\section{$\Xi(1620)$ production in the $K^- p $ scattering process}
\label{sec:MS}
\subsection{Molecular structure of $\Xi(1620)$}

In the molecular frame, $\Xi(1620)$ is considered as a $S$-wave  $\bar{K}\Lambda$ bound state. In the present work, we employ the effective Lagrangian approach to describe the interaction of $\Xi(1620)$ with its components, which is,
\begin{eqnarray}
	\mathcal{L}_{\Xi^\prime K\Lambda} = ig_{\Xi^\prime \Lambda K} \bar{\Xi}^\prime K \Lambda
\end{eqnarray}  
where $\Xi^\prime$ refers to $\Xi(1620)$, and the coupling constant $g_{\Xi^\prime K \Lambda}$ could be determined by the compositeness condition~\cite{Weinberg:1965zz, Baru:2003qq}, which is,
\begin{eqnarray}
	g_{\Xi^\prime K \Lambda}^2
	=\frac{4 \pi} {4 M_{\Xi^\prime} m_{\Lambda}} \frac{(m_K + m_\Lambda)^{5/2}} {(m_K m_\Lambda)^{1/2}} \sqrt{32 \epsilon}
\end{eqnarray}
with $\epsilon=m_{K}+m_{\Lambda}-m_{\Xi^\prime}$ to be the binding energy of $\Xi(1620)$. In the present work, we take $\epsilon = 6~\mathrm{MeV}$, and consequently, the coupling constant is estimated to be $g_{\Xi^\prime K \Lambda }=1.85$.
	
\begin{figure}[t]
	\begin{tabular}{ccc}
		\centering
		\includegraphics[width=65mm]{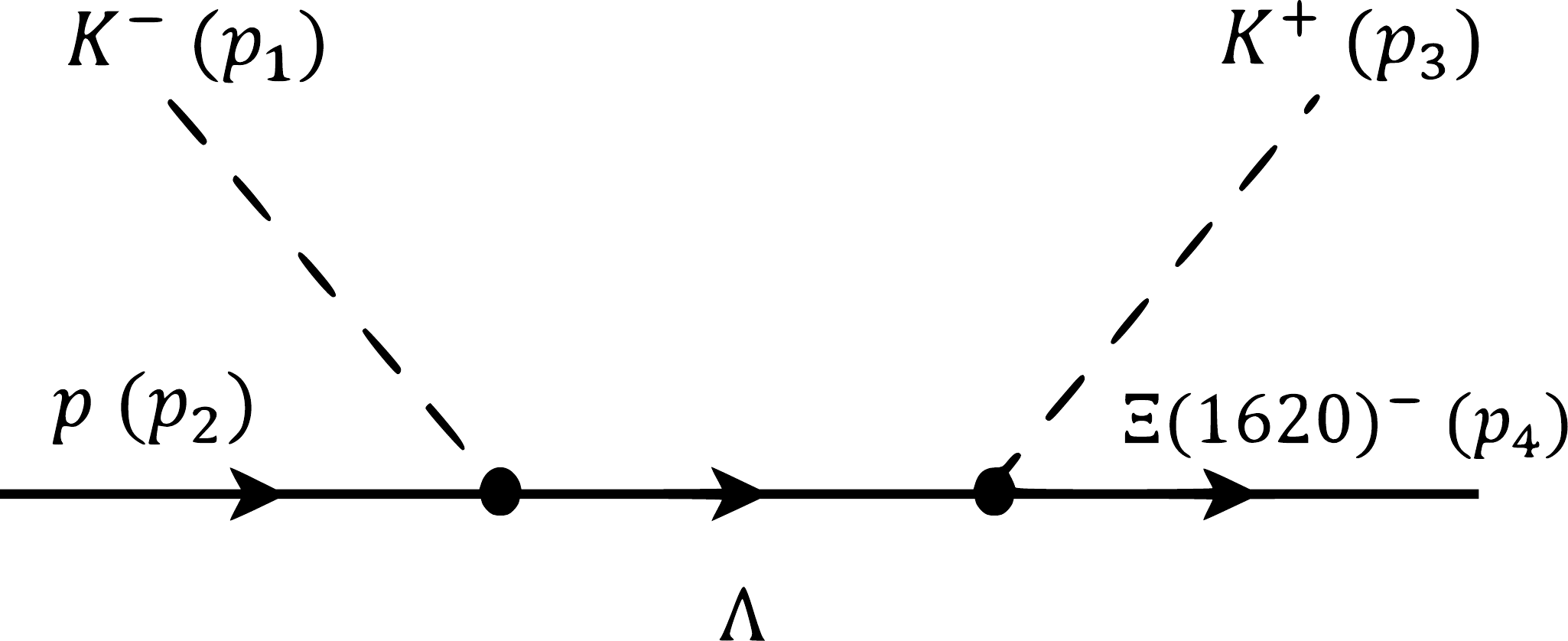}&\\
		{\large(a)} &\\
		\includegraphics[width=65mm]{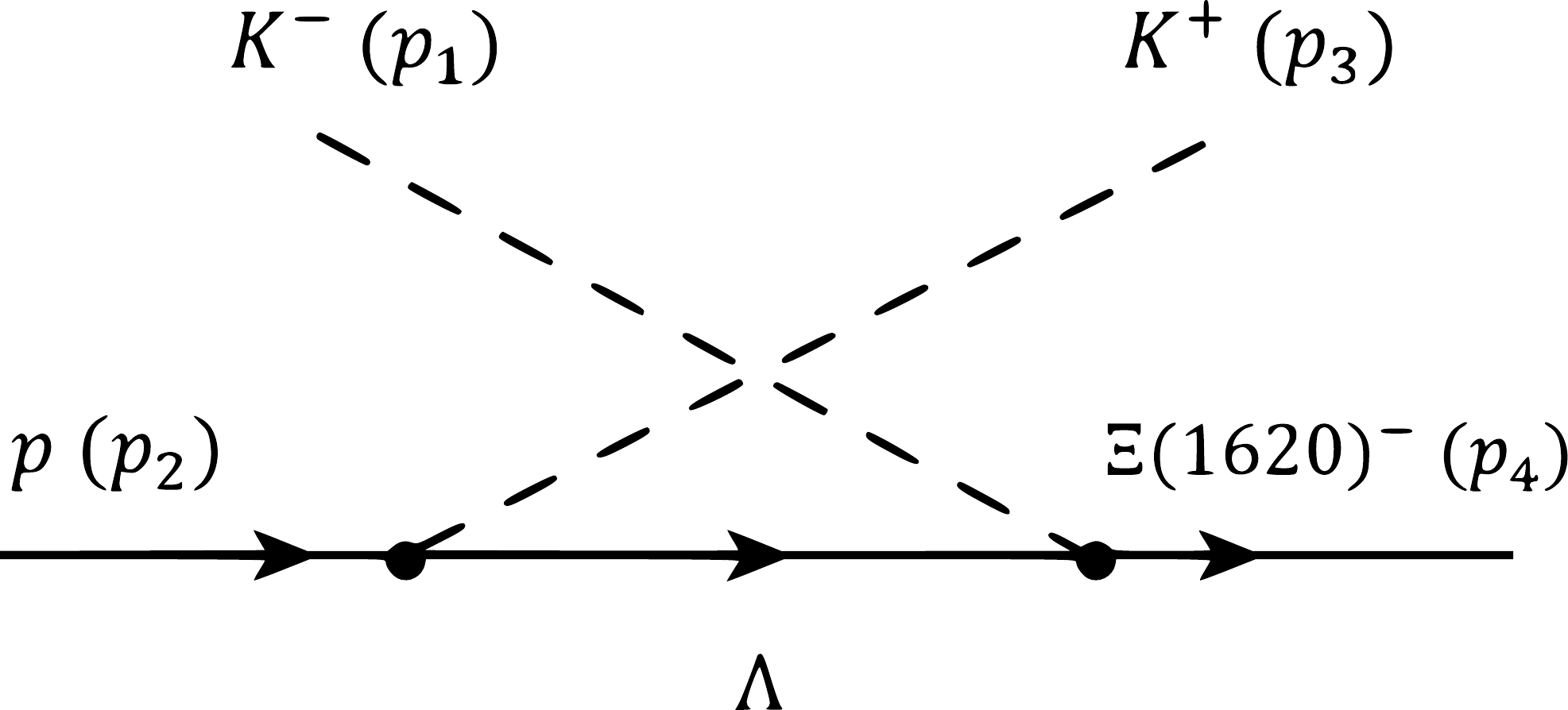}&\\
		\\
		 \large {(b)}& \\
	\end{tabular}
	\caption{Diagrams contributing to the process of $K^- p\rightarrow \Xi(1620)^- K^+$, 
	Diagrams (a) and (b) correspond to the $s$- and $u$-channel contributions, respectively.}\label{Fig:Mech1}
\end{figure}

\subsection{Cross Sections for $K^- p \to \Xi(1620)^- K^+$}
\label{sec:Csf}

In the molecular scenario, $\Xi(1620)$ dominantly couples to its components $\bar{K} \Lambda$.  The diagrams contributing to $K^- p\to K^+ \Xi(1620)$ are listed in Fig~\ref{Fig:Mech1}, where diagrams (a)	and (b) correspond to the $s$- and $u$-channel contributions, respectively. In the present work, we employ the effective Lagrangian approach to depict the hadron interactions. Besides the $\Xi(1620) \Lambda K$ interaction, the effective Lagrangian for $\Lambda NK$ is also need, which is~\cite{Jackson:2015dva}, 
\begin{eqnarray}
	\mathcal{L}_{\Lambda NK}=-ig_{\Lambda NK} \bar{\Lambda}\gamma^5\bar{K} N+H.c. , 
\end{eqnarray}
with the coupling constant $g_{\Lambda N K}=-13.24$~\cite{Jackson:2015dva}.

With above effective Lagrangians, the amplitudes corresponds to Fig.~\ref{Fig:Mech1}-(a) and Fig.~\ref{Fig:Mech1}-(b) read,
\begin{eqnarray}
	\mathcal{M}_{2s} &=& \bar{u}(p_4,m_4) \Big[i g_{\Xi^\prime \Lambda K_c}\Big] \Big[\mathcal{S}^{1/2}  (p_1+p_2,m_\Lambda,\Gamma_{\Lambda})\Big] \nonumber
	\\ &\times &\Big[-i g_{\Lambda NK}\gamma^5\Big]  u(p_2,m_2) F\left((p_1+p_2)^2,m^2 _{\Lambda}, \Lambda_r^2\right)^2 \nonumber\\
\mathcal{M}_{2u} &=& \bar{u}(p_4,m_4) \Big[ i g_{\Xi^\prime \Lambda  K}\Big] \Big[\mathcal{S}^{1/2} (p_2-p_3,m_\Lambda,\Gamma_{\Lambda}) \Big]   \nonumber\\ &\times & \Big[-i g_{\Lambda NK}\gamma^5)\Big]u(p_2,m_2)  F\left((p_2-p_3)^2,m^2 _{\Lambda}, \Lambda_r^2\right)^2, \ \ 
\end{eqnarray}
where $\mathcal{S}(k)$ is the propagator of $\Lambda$ baryon with momentum $k$, and its concrete expression is 
\begin{eqnarray}
\mathcal{S}^{1/2}(k,m,\Gamma) =\frac{\slash\!\!\!k+m_{\Lambda}} {k^2 _{\Lambda} - m^2_{\Lambda} +i m_{\Lambda} \Gamma_{\Lambda}}
\end{eqnarray}
In the above amplitudes, a form factor was introduced to depict the inner structure of the involved hadrons in each vertex, and the concrete form of the form factori is~\cite{Jackson:2015dva},
\begin{eqnarray}
	F(p,m,\Lambda_r)
	=\frac{\Lambda^4 _r} {\Lambda^4 _r +( p^2-m^2)^2} \label{Eq:Fpm}
\end{eqnarray}
Where $p$ and $m$ correspond to the four-momentum and mass of the exchange baryon, while $\Lambda_r$ is a phenomenological model parameter, which will be discussed in the following section.

The total amplitude of $K^-p \to K^+  \Xi(1620)$ is, 
\begin{eqnarray}
	\mathcal{M}_2^{\mathrm{Tot}} =\mathcal{M}_{2s} +\mathcal{M}_{2u},
\end{eqnarray} 
and with this amplitude, the cross sections for $K^- p\rightarrow \Xi(1620)^- K^+$ can be obtained as, 
\begin{eqnarray}
	\frac{d{\sigma}} {d \cos\theta}
	=\frac{1} {32 \pi s} \frac{|\vec{p}_f|} {|\vec{p}_i|}  \left(\frac{1} {2} \left|\overline{\mathcal{M}^{\mathrm{Tot}}_{2}} \right|^2\right),\label{Eq:CS}
\end{eqnarray}
where $s$ and $\theta$ refer for the square of the center of mass energy and the scattering angle, which is the angle of outgoing $\Xi(1620)$ and the kaon beam direction in the center direction. The $\vec{p_f}$ and $\vec{p_i}$ stand for three momenta of the final $\Xi(1620)$ and the initial kaon beam in the center of mass system.

\subsection{Cross Sections for $K^- p \to \Xi^- \pi^0 K^+$}
	\label{sec:Cs}
\begin{figure}[t]
		\begin{tabular}{ccc}
			\centering
			\includegraphics[width=70mm]{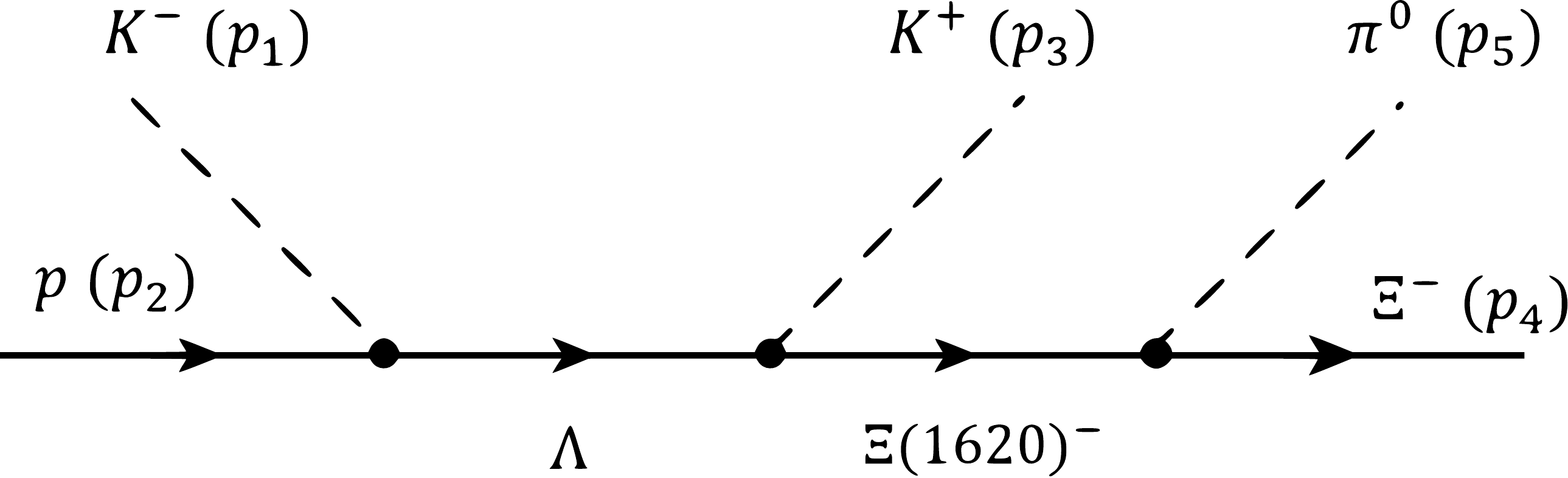}&\\ \\
			{\large(a)}\\ 
			\\
			\includegraphics[width=70mm]{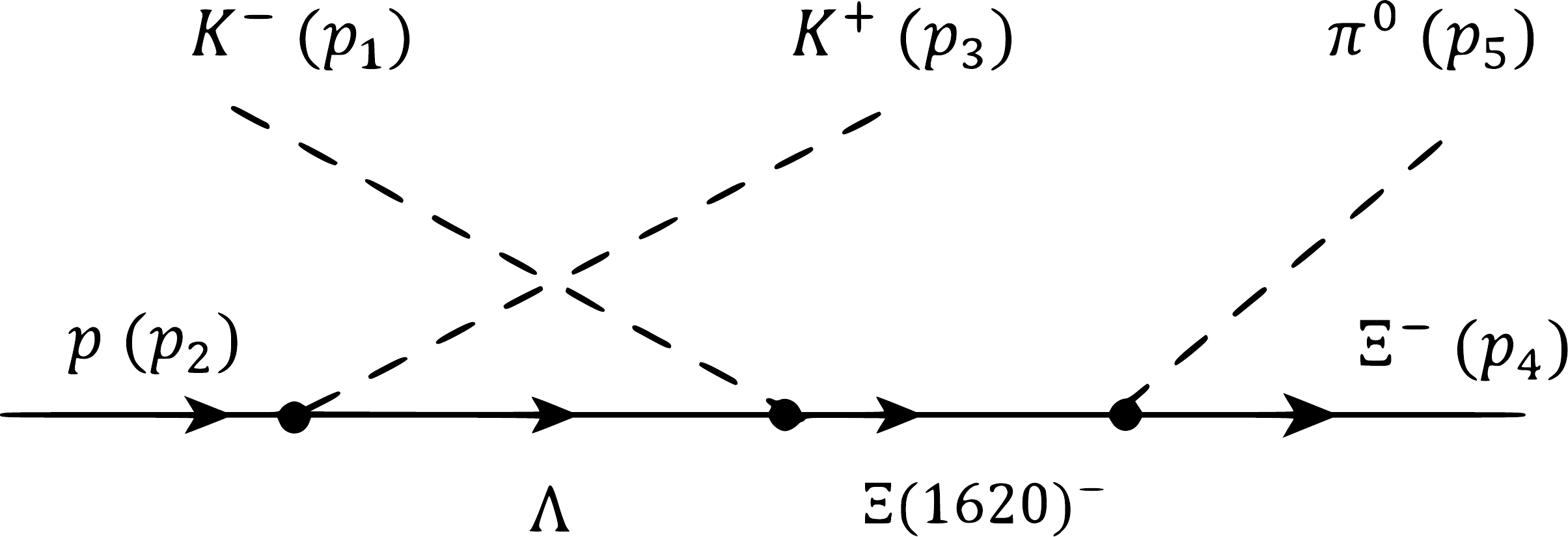}&\\
			{\large(b)}\\
		\end{tabular}
		\caption{Diagrams contributing to the process of $K^- p\rightarrow \Xi^- \pi^0K^+$, Figure (a) corresponds to the s-channel of process, and figure (b) corresponds to the u-channel of process.}\label{Fig:Mech2}
	\end{figure}

It should be noted that the only Okubo-Zweig-Iizuka allowed two body decay of $\Xi(1620)$ is $\Xi\pi$. Thus, one can expect to detect the signal of $\Xi(1620)$ in the invariant mass distributions of $\Xi\pi$ of process $K^- p \to \Xi^- \pi^0 K^+$, thus, in the following subsection, we will estimate the cross sections for $K^- p \to \Xi^- \pi^0 K^+$, where $\Xi^- \pi^0$ come from the decay of $\Xi(1620)^-$. The diagrams contributing to the process $K^- p \to \Xi^- \pi^0 K^+$ are presented in Fig.~\ref{Fig:Mech2}, where diagrams (a) and (b) denote the s channel and u channel contributions, respectively. To estimate the diagrams in Fig.~\ref{Fig:Mech2}, the effective Lagrangian of $\Xi(1620)\Xi \pi$ is needed, which is, 
	\begin{eqnarray}
	\mathcal{L}_{\Xi^\prime \Xi \pi}=ig_{\Xi^\prime \Xi \pi} \bar{\Xi}(\mathbf{\tau}\cdot \mathbf{\pi}) \Xi^\prime+H.c.
\end{eqnarray}
The coupling constant $g_{\Xi^\prime \Xi \pi}$ could be estimated from the partial width of $\Xi(1620)\to \Xi \pi$.

 For the coupling constant $g_{\Lambda\Xi(1620)K}$, we consider the decay process $\Xi(1620) \rightarrow \Xi \pi$, which combined with the corresponding decay width $\Gamma =59.9 \mathrm{MeV}$ \cite{Wang:2019krq} and the effective Lagrangian for the $\mathcal{L}^{1/2(\pm)}_{\Xi(1620) \Xi \pi}$, the result can be calculated as $g_{\Xi(1620)\Xi\pi}=1.01$; 

With the above effective Lagrangian, the amplitudes corresponding to the diagrams in Fig.~\ref{Fig:Mech2} read,
\begin{eqnarray}
	\mathcal{M}_{3s} &=& \bar{u}(p_4,m_4) \Big[i g_{\Xi^\prime \Xi \pi}\Big] \Big[S^{1/2} (p_4+p_5,m_{\Xi^\prime},\Gamma_{\Xi^\prime})\Big] \nonumber
	\\ &\times & \Big[i g_{\Xi^\prime \Lambda K}\Big] 
	\Big[S^{1/2} (p_1+p_2,m_\Lambda,\Gamma_{\Lambda})\Big] \Big[-i g_{\Lambda NK}\gamma^5\Big] u(p_2,m_2) \nonumber
	\\ &\times & F(p^2 _s,m^2 _{\Lambda})  F(p^2 _s,m^2 _{\Lambda}) F(p^2 _{\Xi^\prime},m^2 _{\Xi^\prime},\Lambda_r^2)\nonumber\\
	\mathcal{M}_{3u} &=& \bar{u}(p_4,m_4) \Big[ i g_{\Xi^\prime \Xi \pi}\Big] \Big[S^{1/2}  (p_4+p_5),m_{\Xi^\prime},\Gamma_{\Xi^\prime}\Big] \nonumber\\
	&\times &\Big[ g_{\Xi^\prime \Lambda   K }\Big] 
	\Big[S^{1/2} (p_2-p_3,m_\Lambda,\Gamma_{\Lambda})\Big] \Big[-i g_{\Lambda NK}\gamma^5\Big]u(p_2,m_2) \nonumber
	\\ &\times & F(p^2 _u,m^2 _{\Lambda})  F(p^2 _u,m^2 _{\Lambda}) F(p^2 _{\Xi^\prime},m^2 _{\Xi^\prime}).
\end{eqnarray}
where the form factor is the same as the one in Eq.~\eqref{Eq:Fpm}.

Then the total amplitude for $K^- p\to K^+ \pi^- \Xi^{-}$ is, 
\begin{eqnarray}
	\mathcal{M}_{3}^{\mathrm{Tot}} =\mathcal{M}_{3s}+\mathcal{M}_{3u},
\end{eqnarray}
and the corss sections for the $K^- p\rightarrow \Xi^- \pi^0K^+$ is 
\begin{eqnarray}
	d{\sigma}
	=\frac{1} {8(2 \pi)^4} \frac{1} {	\Phi} \left|\overline{\mathcal{M}_{3}^{\mathrm{Tot}}}\right|^2 d p^0 _5 d p^0 _3 d\cos\theta d \eta
\end{eqnarray}
with the flux factor $\Phi=4|{\vec{p_1}}|\sqrt{s}$, and $\vec{p_1}$ and $\sqrt{s}$ represent the three momentum of the initial $K^-$ and the center of mass energy, respectively. $p^0_3$ and $p^0_5$ are the energy of the outgoing $K^+$ and $\pi^0$, respectively.

\section{Numerical Results and discussions }
\label{sec:NR}

\begin{figure}[htb]
		\includegraphics[width=85mm]{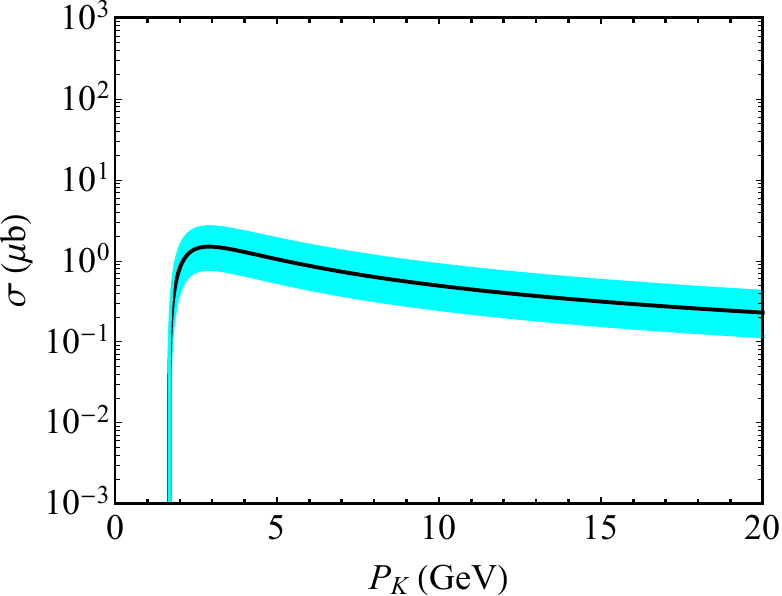}
	\caption{The cross sections for $K^- p\rightarrow \Xi(1620)^- K^+$ depending on the momentum of the incident kaon beam.} \label{Fig:CS2}
\end{figure}

\subsection{Cross sections and differential cross sections for $K^-p \to K^+ \Xi(1620)^-$}

Before the we estimate the cross sections of the discussed processes, the model parameter $\Lambda_r$ introduced by the form factor should be discussed. In Ref.~\cite{Jackson:2015dva}, the process  $\bar{K}p \to K \Xi $ were investigated with an effective Lagrangian approach. The estimated cross sections could be well reproduced with $\Lambda_r=0.90$ GeV. In this work, the considered processes are very similar to the one in Ref.~\cite{Jackson:2015dva}, thus the model parameters $\Lambda_r$ in the present estimations should be close to the one used in Ref.~\cite{Jackson:2015dva}, and in the present work, we vary $\Lambda_r$ from 0.87 to 0.97 GeV to check the parameter dependences of the cross sections for the considered processes.

With the above preparation, the cross sections for the $K^- p\rightarrow \Xi(1620)^- K^+$ process could evaluated. The cross sections depending on the momentum of the incident kaon are presented in Fig.~\ref{Fig:CS2}, where the black curve is estimated with $\Lambda_r=0.92$ GeV, while the cyan band indicates the uncertainties resulted from the variation of the model parameter $\Lambda_r$. One can find the cross sections increase sharply near the threshold of $K^+ \Xi(1620)^-$, and then the cross sections decrease slowly with the increasing of the momentum of the kaon beam. In particular, the cross section is $(1.49^{+1.12}_{-0.70})~\mathrm{\mu b}$ at $P_K=2.87$ GeV.

\begin{figure}[t]
		\centering
		\includegraphics[width=85mm]{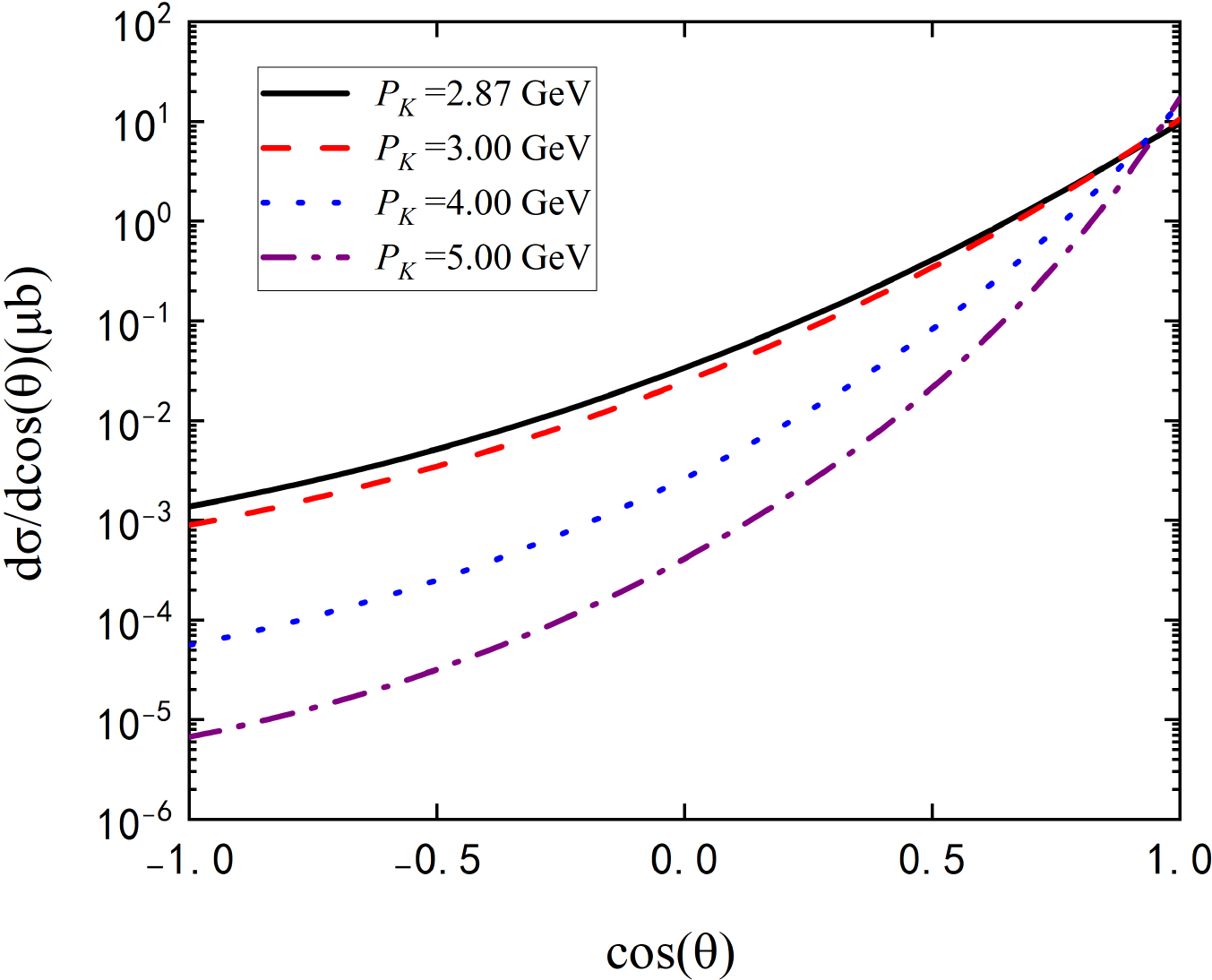}
	\caption{The differential cross section for  $K^- p\rightarrow \Xi(1620)^- K^+$ depending on $\cos\theta$ for several typical momenta of the incident kaon beam.} \label{Fig:DCS2}
\end{figure}

 Besides the cross sections, we also estimate the different cross sections for $K^-p \to K^+ \Xi(1620)^-$ depending on $\cos \theta$, where $\theta$ is the angle between the outgoing $\Xi(1620)$ and the beam direction. Here, we take the model parameter $\Lambda_r=0.92$ GeV, and the momenta of the incident kaon beam as $2.87$, $3.00$, $4.00$ and $5.00$ GeV, respectively. The $\cos \theta $ dependences of the the differential cross sections are presented in Fig.~\ref{Fig:DCS2}. From the figure one should notice that the differential cross sections reach the maximum at the forward angle limit. As $P_K$ increases, more $\Xi(1630)$ are concentrated in the forward angle area.


\subsection{Cross sections for $K^- p \to K^+ \Xi^- \pi^0$}

\begin{figure}[t]
		\centering
		\includegraphics[width=85mm]{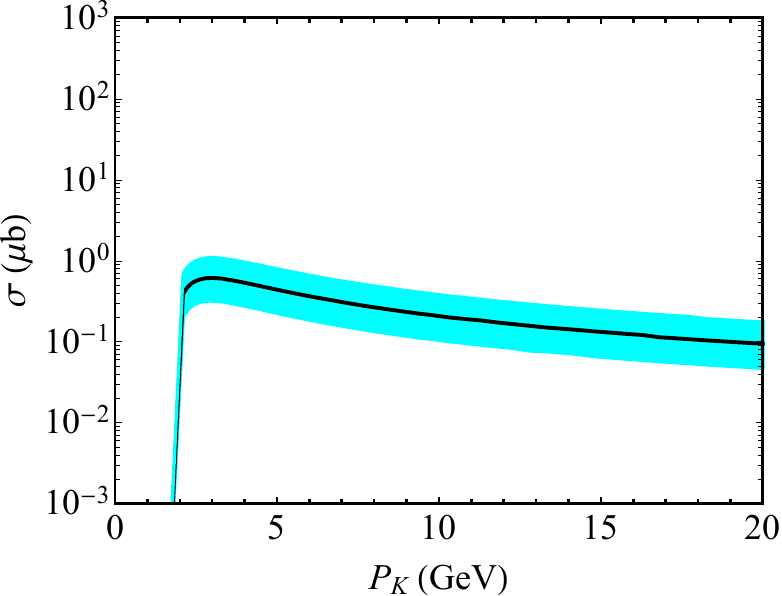}
	\caption{The cross sections for the process $K^- p\rightarrow \Xi^- \pi^0K^+$ depending on the momentum of the incident kaon beam.} \label{Fig:CS3}
\end{figure}

In Fig.~\ref{Fig:CS3}, we present the cross sections for the process $K^- p\rightarrow \Xi^- \pi^0 K^+$, where $\Xi^- \pi^0$ are the daughter particles of $\Xi(1620)$. The $P_K$ dependences of the cross sections are very similar to that of the cross sections for $K^- p\to K^+ \Xi(1620)^-$. In particular, the cross sections increase rapidly near the threshold of $\Xi^- \pi^0 K^+$, and the smoothly decrease with the increasing of $P_K$. In the considered model parameter range, we find the cross sections for $K^- p \to \Xi^- \pi^0 K^+$ is estimated to be $(0.61^{+0.47}_{-0.29})~\mathrm{\mu b}$ at $P_K=2.87$ GeV, which is safely under the upper limit of the experimental measurement.

\section{Summary}
\label{sec:Summary}	

The $\Xi(1620)$ state has been studied continuously over the years. Recently, the experimental observation of the double strange baryon $\Xi(1620)$ from the Belle Collaboration is of great significance to its research, through which we are able to obtain precise mass and width data for the $\Xi(1620)$ particle. The observed mass of $\Xi(1620)$ are very close the the threshold of $\bar{K} \Lambda $, which indicates that $\Xi(1620)$ could be a good candidate molecular state composed of $\bar{K}\Lambda$.

Besides the resonance parameter, the production process could also provide some crucial information of the inner structure of $\Xi(1620)$. In the present work, the production process $K^- p\to K^+\Xi(1620)^-$ are investigated. The cross sections and the differential cross sections for $K^- p\to K^+\Xi(1620)^-$ are evaluated. Our estimations indicate that the cross sections for $K^- p \to K^+ \Xi(1620)^-$ increase sharply near the threshold of $K^+ \Xi(1620)^-$, and then decrease slowly with the increasing of $P_K$. In particular, the cross section is estimated to be  $(1.49^{+1.12}_{-0.70})~\mathrm{\mu b}$ at $P_K=2.87$ GeV. In addition, we find the differential cross sections reach the maximum at the forward angle limit.

Consider the kinematics limit, $\Xi(1620)$ should dominantly decay into $\Xi \pi $, thus in the present work, we also estimate the cross sections for $K^- p\rightarrow \Xi^- \pi^0K^+$, where $\Xi^- \pi^0$ come from the decay of $\Xi(1620)$. The upper limit of  cross sections for $K^- p\rightarrow \Xi^- \pi^0K^+$ was determined to be $1.2~\mathrm{\mu b}$ at $P_K=2.87~$MeV, and the cross section for $K^- p\rightarrow \Xi^- \pi^0K^+$ is estimated to be $(0.61^{+0.47}_{-0.29})~\mathrm{\mu b}$ at $P_K=2.87$ GeV, which is safely under the upper limit of the experimental measurement.

		\section*{ACKNOWLEDGMENTS}
	This work is supported by the National Natural Science Foundation of China under the Grant 12175037 and 12335001.

\end{document}